\documentclass[notitlepage]{report}
\usepackage{titling}
\usepackage[caption2]{ccaption}
\usepackage{xr-hyper}
\usepackage{hyperref}
\usepackage{nameref}
\usepackage{zref-xr}
\zxrsetup{toltxlabel}

\usepackage[utf8]{inputenc}
\usepackage{pdflscape}
\usepackage{amssymb,amsbsy,color,amsmath}
\usepackage{graphicx}
\usepackage{dcolumn}
\usepackage{pgfplotstable}
\usepackage{filecontents}
\usepackage{longtable}
\usepackage{authblk}
\usepackage{hyperref}

\usepackage{subcaption}

\usepackage{authblk}
\usepackage{pdfpages}
\pdfoutput=1

\begin{document}
%

\title{Linguistic evolution driven by network heterogeneity and the Turing mechanism}
\newcommand{\RochesterP}{Department of Physics \& Astronomy, University of Rochester, Rochester, NY 14607, USA}
\newcommand{\SantiagoN}{Group of Nonlinear Physics. University of Santiago de Compostela, 15782 Santiago de Compostela, Spain}
\newcommand{\SantiagoW}{Dept. de F\'isica Aplicada. University of Santiago de Compostela, 15782 Santiago de Compostela, Spain}
\newcommand{\Korea}{Graduate School of Culture Technology, Korea Advanced Institute of Science and Technology, Daejon, 305-701, Korea}

\author[1]{Sayat Mimar}
\author[2]{Mariamo Mussa Juane}
\author[3]{Jorge Mira}
\author[4]{Juyong Park}
\author[2]{Alberto P. Mu\~nuzuri}
\author[1]{Gourab Ghoshal}

\affil[1]{\RochesterP}
\affil[2]{\SantiagoN}
\affil[3]{\SantiagoW}
\affil[4]{\Korea}


\date{}
\makeatletter
\newcommand*{\toccontents}{\@starttoc{toc}}
\makeatother

\maketitle
\toccontents

\newpage

\phantomsection
\addcontentsline{toc}{chapter}{Manuscript}

\maketitle
\renewcommand{\thesection}{\arabic{section}}
\renewcommand{\thefigure}{\arabic{figure}}
\renewcommand{\thetable}{\arabic{table}} 
\renewcommand{\theequation}{\arabic{equation}}

\begin{abstract}
Given the rapidly evolving landscape of linguistic prevalence, whereby a majority of the world's existing languages are dying out in favor of the adoption of a comparatively fewer set of languages, the factors behind this phenomenon has been the subject of vigorous research. The majority of approaches investigate the temporal evolution of two competing languages in the form of differential equations describing their behavior at large scale. In contrast, relatively few consider the spatial dimension of the problem. Furthermore while much attention has focused on the phenomena of language shift---the adoption of majority languages in lieu of minority ones---relatively less light has been shed on linguistic coexistence, where two or more languages persist in a geographically contiguous region. Here, we study the geographical component of language spread on a discrete medium to monitor the dispersal of language species at a microscopic level. Language dynamics is modeled through a reaction-diffusion system that occurs on a heterogeneous network of contacts based on population flows between urban centers. We show that our framework accurately reproduces empirical linguistic trends driven by a combination of the Turing instability, a mechanism for spontaneous pattern-formation applicable to many natural systems, the heterogeneity of the contact network, and the asymmetries in how people perceive the status of a language. We demonstrate the robustness of our formulation on two datasets corresponding to linguistic coexistence in northern Spain and southern Austria.

\end{abstract}

\section{Introduction}
Language is the center of human activity and has served as the fundamental mode of communication since the dawn of human civilization. While there currently exists roughly 6000 languages differing structurally in terms of grammar and vocabulary~\cite{Kandler2017}, they all evolve dynamically through human interactions, that are shaped by economic, political, geographic and cultural factors ~\cite{Tsunoda2006,Nettle1998,Castellano2009,Ball2002}. 
\par

A rather unfortunate outcome of such evolution is the replacement of vernacular tongues, spoken by a minority of the population, with that spoken by the dominant majority. Indeed it is estimated that 90\% of existing languages will go extinct by the end of the century~\cite{Kandler2017,Unesco}, leading to a huge loss in cultural diversity, given the inextricable links between speech and customs. One of the first mathematical models that accurately reproduced such ``language-death" was proposed by Abrams and Strogatz (AS)~\cite{Abrams2003}. In their formulation, two languages compete, with the attractiveness of each of the species being determined by its perceived status amongst the population. As long as the symmetry between the perceived status is broken, the model necessarily predicts a single hegemonic language adopted by the entire population. Indeed, the model successfully accounts for the decline of 42 real-world minority languages in contact with hegemonic counterparts.  The formulation however fails to account for those cases where languages coexist in a geographically contiguous region.   
\par
To account for this limitation, refinements were made to the AS model by Mira and Paredes \cite{mira2005interlinguistic, Seoane2017} incorporating bilingualism by introducing an inter-linguistic similarity parameter. An important example of this occurs in the northwestern part of Spain in Galicia, where both Galician and Spanish are spoken. Their model analyzes the temporal evolution of these languages demonstrating the existence of a stable coexistence given enough similarity between the languages. 

\par
Both formulations and other related ones~\cite{mira2011importance,Zhang2013,Isern2014} focus on the temporal aspect of language evolution at a macroscopic scale, while ignoring local dynamics on the space where subpopulations of competing tongues reside. To address this, other approaches incorporate the geographic component into reaction-diffusion equations~\cite{Cooper2013, Cooper2013b, Ghoshal2016} simulating the  dispersal of speakers in a continuous domain~\cite{Kandler2009,Yun2016,Kandler2010,Patriarca2009}. While the approach is reasonable, it fails when geographic regions representing speakers of a common language are no longer contiguous, and thus there is no meaningful diffusive front.  Furthermore, the approach is unable to provide spatially detailed description of language spread and retreat. 

\par
An agent-based probabilistic model proposed in~\cite{Prochazka2017}, supported with detailed empirical data from southern Austria shed light on the constituents of language dynamics at a microscopic level. The region, where speakers of German and Slovenian live, is partitioned into quadratic grids where each cell represents an area of one square kilometer. To determine the probability of speaking one of the languages in a given year, the model uses the number of speakers of each language in the preceding year for every cell and their interaction with speakers with surrounding cells, hence accounting only for short-range connections. Although this fine-grained model is able to successfully determine the temporal evolution of the two languages and generate satisfactory results for geographic distribution of the subpopulations, it has limitations in terms of generalization. The historic and elaborate dataset  for the number of speakers covering the entire region are from the periods 1880-1910 and 1971-2001, and such detailed records are not so easily available for other parts of the world. 
\par
A recent work~\cite{MussaJuane2019} proposed a district level mean field network with uniform weights for Galicia where two competing languages are spread across 20 districts of the region with different prestige values, to combine internal complexity of each location with influence produced by its neighbors. This approach explains how the interplay between urban and rural dynamics leads to competition in language shift. The framework however requires fine-grained detail on a plethora of parameters to study the sociolinguistic dynamics across the region.

\par 
While the models described thus far, reproduce, to varying degrees of accuracy, the evolution of the observed linguistic trends (to the extent that such data is available) they are formulated in a fashion that makes it difficult to disentangle the effects of the various mechanisms governing linguistic evolution. Additionally while focusing either on short range interactions or at a macroscopic scale, none of the models consider the effect of human mobility, a rather important ingredient in understanding the dynamics of socioeconomic systems~\cite{Lee_2017, Kirkley_2018, Barbosa2018}. Here we propose a coarse-grained model of language dynamics that seeks to uncover the minimal mechanisms that reproduce the observed linguistic trends. We set up our model in such a way that we can interrogate the effect of each of the constituent mechanisms. It is important to note that our goal is not to reproduce exactly the number of speakers of a given language, but rather, sacrificing specificity and erring on the side of generalizability, we focus on the qualitative trends. 

Our model consists of primarily three ingredients. We first discretize the space on which linguistic interactions occur by representing towns as nodes and connections between them as edges, incorporating population flows at microscopic level. This geographic network extends the work of~\cite{Prochazka2017} by combining both short-range and long-range connections that are not present in the agent-based model and are essential to describe global population interactions. The edges are weighted by a gravity-like relation~\cite{Barbosa2018,Zipf1946, Simini2012,Pan_2013_UrbanScaling} which is the simplest parameter free model to calculate mobility flows between two communities, while accounting for their geographic separation. Second, the evolution of each language is characterized by reaction-diffusion equations for two competing languages whose dynamics is described by the Lotka-Volterra model, previously used to model linguistic coexistence~\cite{Pinasco2006, Kandler2008}. As opposed to wavefront propagation in continuous space which requires a contiguous region, reaction diffusion on networks prevents the isolation of language islands. The contact network enables interactions along weighted edges such that each node communicates with all of its neighbors. Spatial linguistic patterns emerge on the geographical network through the Turing Mechanism,  an exemplar of pattern formation~\cite{Castets1990, Ouyang1991, Sayama2003} that relates to many observed natural phenomena in continuous~\cite{Turing1952} and discrete media~\cite{Nakao2010, Mimar2019}. The final ingredient in our model is the status perception of each language and its corresponding degree of spatial correlation.
\par
We test our model in two different regions of the world with linguistic coexistence, Galicia, in northwestern Spain, where Galician and Spanish are spoken, as well as Carinthia in southern Austria, where one finds both Slovenian and German speakers. In both cases, we find excellent agreement with qualitative trends, demonstrating that in addition to the specific model of linguistic competition one must also account for the geographic network on which the dynamics take place, as well as asymmetries in  linguistic status perception.

 \begin{figure}[t!]
	\captionsetup{justification=raggedright, font=small, labelfont=bf}
	
	\includegraphics[width =1.0\columnwidth]{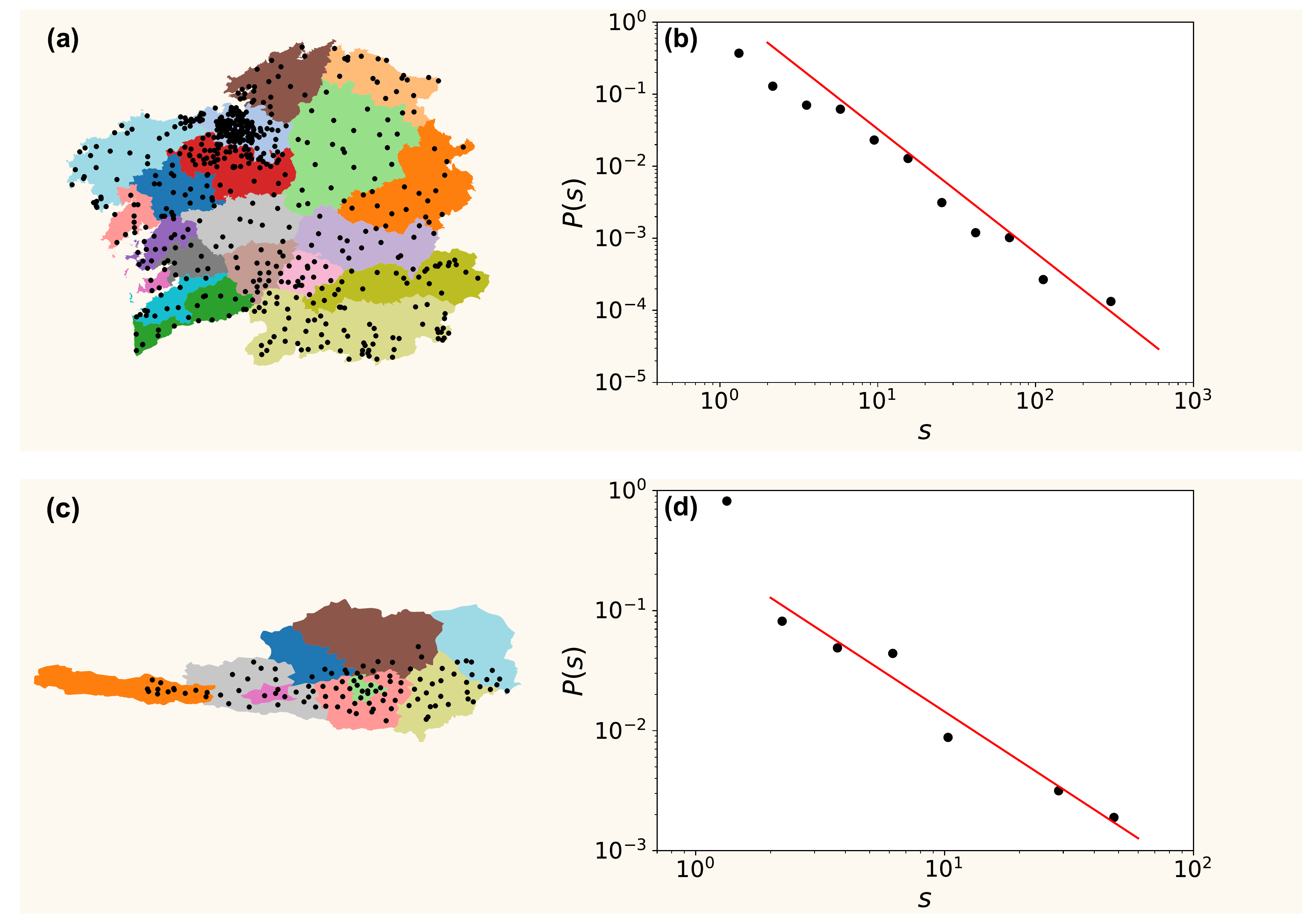}
	
	\caption{\textbf{(a)} 20 districts of Galicia, each represented by a a different color. The points (nodes) correspond to the 550 cities and towns in the region.  \textbf{(c)} The corresponding map for Southern Carinthia comprising of 9 districts and 112 cities. \textbf{(b)} and \textbf{(d)} Distribution of edge-weights $s_i = \sum_j W_{ij}$, for the geographic networks in Galicia and Southern Carinthia with logarithmic binning. The tails of both distributions has the form $P(s) \sim s^{-\beta}$. The exponents are extracted using maximum likelihood estimation and are $\beta^G \approx 1.72$ in \textbf{(b)} and $\beta^C \approx 1.35$ in \textbf{(d)}.}
\label{fig:rawmap}
\end{figure}

 \section{Data}
 
\begin{figure}[t!]
	\centering
	
	\captionsetup{justification=raggedright, font=small, labelfont=bf}
	\includegraphics[width =0.9\columnwidth]{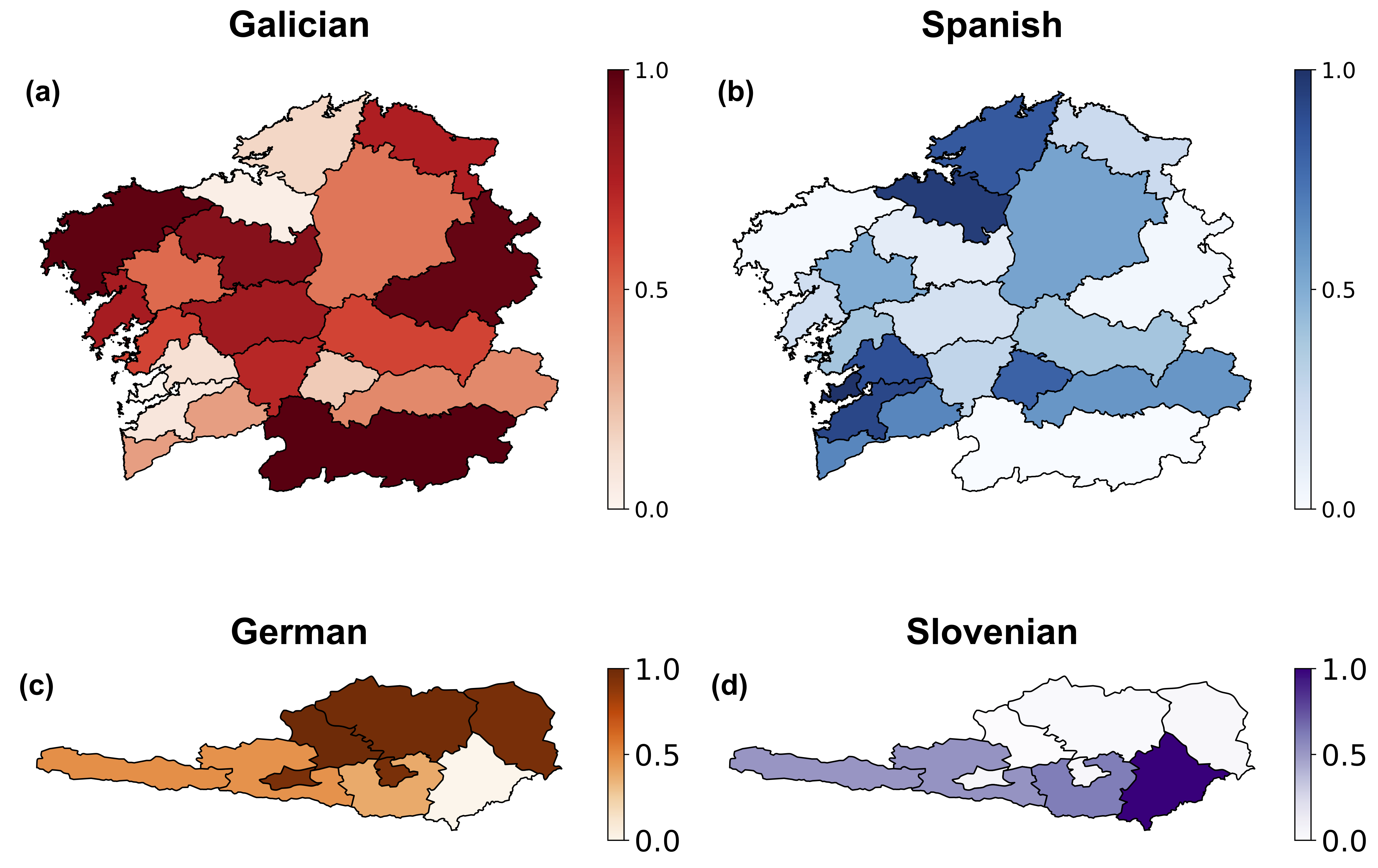}

	\caption{Spatial distribution of linguistic prevalence for Galician \textbf{(a)} and Spanish \textbf{(b)} speakers in the region of Galicia. Regions are colored according to the fraction of speakers in that particular district. The same map for German \textbf{(c)} and Slovenian \textbf{(d)}  speakers in Southern Carinthia. In both regions, languages are geographically distributed in a complementary fashion; regions with the most abundant speakers of a particular language correspond to the lowest amount of speakers of the other language.}
	\label{fig:langmap}
	\end{figure}

In our analysis we make use of datasets corresponding to two independent parts of Europe where populations speak at least two languages. The first comes from the Autonomous Region of Galicia in northwestern Spain, where Galician (a Romance language similar to Portuguese) and Castillian (Spanish) are co-official. The dataset includes information about the fraction of Galician and Spanish speakers in 20 districts of the region consisting of 550 cities \cite{Galicia-fraction}. In Fig.~\ref{fig:rawmap}\textbf{a}, we illustrate the different districts with distinct colors for each region. The cities are represented as black points. The linguistic distributions according to the census data are shown in Fig.~\ref{fig:langmap}\textbf{a,b} for Galician and Spanish, where the colors represent the fraction of speakers of a particular language in each district. The figure indicates that the two languages are geographically distributed in a complementary fashion; regions with the most abundant Galician speakers correspond to the lowest amount of Spanish speakers and vice versa. 

Our second dataset corresponds to Slovenian and German speakers in Southern Carinthia, Austria in the year 1910 across 9 districts and 112 cities. The data available in digitized form~\cite{Carinthia-data} consists of the fraction of the population speaking either language at the level of cities. The region with each district corresponding to a different color, with the cities represented as points is shown in Fig.~\ref{fig:rawmap}\textbf{c}. In Fig.~\ref{fig:langmap}\textbf{b} we plot the spatial distribution of prevalence for the year 1910, finding a similar trend to that seen for Galicia; regions consisting of large number of Slovenian speakers (primarily near the Austrian-Slovenian border) consist of few German speakers, with the opposite also being true. 


\par


 \section{Model}

\subsection{Reaction-diffusion equations} 
We model the language competition with the following set of Lotka-Voltera type differential equations, first proposed in~\cite{Pinasco2006,Kandler2008}:
\begin{align}
\frac{du}{dt} &= cuv + \alpha_u u(1 - \frac{u}{S_u}), \nonumber \\
\frac{dv}{dt} &= -cuv + \alpha_v v(1 - \frac{v}{S_v}).
\label{eq:lv}
\end{align}
Here $u(\vec x, t)$ and $v(\vec x, t)$ correspond to the frequency of the population speaking each language, whereas the cross-term $u v$ represents the competition between them with a strength $c$, interpreted as the status of the language; $c > 0 $ indicates that language $u$ has a higher status or attractiveness than language $v$, with the opposite being true for $c <0$. In the absence of competition, the model reduces to logistic growth where $S_u$ and $S_v$ represent the respective carrying capacities of the languages and $\alpha_u$, $\alpha_v$ their natality and mortality rates. For the purposes of our analysis these constants are redundant and without any loss of generality set to 1 except for $c$, which is constrained to $|c| < 1$. In this setting, the fixed points of Eq.~\eqref{eq:lv} are $(u_0,v_0) = (\frac{1 + c}{1 + c^2}, \frac{1 - c}{1 + c^2})$.

\par
In continuous media, the spatial component representing the diffusion of species is introduced via a second-order diffusion coefficient~ \cite{Yun2016,Kandler2008} proportional to $\nabla^2 u$. It's counterpart in discrete media, such as networks, is the Laplacian matrix $L_{ij}$~\cite{Nakao2010,Othmer1971}, defined as: 
\begin{equation}
L_{ij} = W_{ij} - s_i \delta_{ij},
\label{eq:laplace}
\end{equation}
where $W_{ij}$ is a symmetric weighted adjacency matrix and $s_i =\sum_j W_{ij}$ is the weighted degree of node $i$~\cite{Ioannis2007}. To include diffusion on the underlying network one would then add a term of the form $\sum_{j=1}^N L_{ij} u_j$, and a corresponding one for $v$, in a network of $N$ nodes ($i = 1 \dots N$). If one were to consider ordinary diffusion then one need only include a diffusion coefficient $d_u, d_v$ for each of the populations, which represents the diffusing away from populations of higher density to lower density regions. Yet, in the context of competition, one must also consider effects where a minority population under threat from a majority population diffuses away to a different region to avoid extinction~\cite{Vidal-Franco2017, Vanag2009}. Such cross-diffusion can be introduced by corresponding coefficients $a_{uv}, a_{vu} \geq 0$ proportional to the product $uv$. With these refinements, Eq.~\eqref{eq:lv} can be recast thus,

\begin{align}
\frac{du_i}{dt} &= cu_iv_i + u_i(1 - u_i) + \sum_{j=1}^N L_{ij}[(d_u + a_{uv}v_j)u_j], \nonumber \\
\frac{dv_i}{dt} &= -cu_iv_j +  v_i(1 - v_j) + \sum_{j=1}^N L_{ij}[(d_v + a_{vu}u_j)v_j].
\label{eq:lvd}
\end{align}
Note, that in districts where $c>0$, the higher status tongue corresponds to the population $u$, and $v$ tends to move away at a rate $a_{uv} = \gamma$ whereas $u$ remains in the district so $a_{vu} = 0$. Similarly, in districts where $c<0$, $u$ diffuses away and we have $a_{uv} = 0$, $a_{vu} = \gamma$.


\par

\subsection{Choice of network and Turing instability}
Next, we consider the choice of network that best represents the interactions between the populations in different centers. Since we are interested in capturing the effects of population movement, in principle, all cities are accessible to each other through a transportation network, such that all nodes are connected to each other as in a complete graph. Yet the extent of flows between two cities $i,j$ depends on their respective populations $p_i, p_j$ and the distance $d_{ij}$. The simplest choice for the coupling between cities is the gravity model~\cite{Barbosa2018,Zipf1946} with weights
\begin{equation}
W_{ij} = \frac{p_jp_j}{d_{ij}^2}.
\label{eq:gravity}
\end{equation}
Here $W_{ij}$ represents the $i^{th}$ row and $j^{th}$ column of the weight matrix $\textbf{W}$; $p_i$ and $p_j$ are the populations of the corresponding cities and $d_{ij}$ is the geographical distance separating them. Thus greater flows occur either between high population centers or those proximate to each other. In principle the gravity model can be generalized to more complex dependencies on the population and distance~\cite{schneider_1959_gravity} and one can consider related models of mobility such as those based on intervening opportunities~\cite{Simini2012,ortuzar_2011_modeling}, however the version considered here  has been used to accurately described mobility patterns in different contexts~\cite{Jung2008,Noulas2012}, and our results are not overly sensitive to the precise form of Eq.~\eqref{eq:gravity} as long as the networks are heavy-tailed~\cite{Mimar2019}.  In Fig.~\ref{fig:rawmap}\textbf{b,d} we show the strength distribution $P(s)$ of the Galician and Southern Carinthian geographical networks indicating a right-skewed distribution in both cases. A maximum likelihood fit to the tails of the distribution yields  $P(s) \sim s^{-\beta}$ with $\beta^G \approx 1.72$ and $ \beta^C \approx 1.35$. The average weights are $\langle s \rangle^G $ = $6.33$ and $\langle s \rangle^C $ = $1.54$ whereas the variance $\sigma_s = \sqrt{\langle s^2 \rangle  - \langle s \rangle^2 } $ for each network are $\sigma_s^G$ = $29.38$, and $\sigma_s^C$ = $5.47$. 

\par

Equation~\eqref{eq:lvd} with the appropriate parameter values can exhibit Turing structures, i.e. stationary non-homogeneous solutions~\cite{Turing1952}, recently proposed as a mechanism to  explain spatial differentiation in linguistic competition~\cite{Nakao2010,Vidal-Franco2017}. 
A small perturbation to the uniform state triggers the growth of Turing patterns above a critical threshold, corresponding to the ratio of the diffusion constants of the respective linguistic species. The patterns in this context, correspond to distinct populations of nodes differentiated by the relative number of the population speaking a certain language. While in continuous media, perturbations are decomposed into a set of spatial Fourier
modes representing plane waves with different wave-numbers, in networks the analog is the set of eigenvectors $\boldsymbol{\phi}^{(\alpha)}$ of the Laplacian matrix (with associated eigenvalue $\Lambda_{\alpha}$), where $\alpha = 1,\ldots N$ corresponds to the mode~\cite{Othmer1971}. The eigenvalues $\Lambda_{\alpha}$ are sorted in decreasing order  $\Lambda_{1} >\Lambda_{2} \dots>\Lambda_{N}$ and the first eigenvalue is always zero ($\Lambda_{1} = 0$). Introducing small perturbations $(\delta u_i, \delta v_i)$, substituting into Eq.~\eqref{eq:lvd}, and expanding over the set of the Laplacian eigenvectors, the linear growth rate $\lambda_{\alpha}$ for each node is calculated from a polynomial equation of the form  
\begin{equation}
\lambda_{\alpha}^2 + b(\Lambda_{\alpha})\lambda_{\alpha} + c(\Lambda_{\alpha}) = 0,
\label{eq:charac}
\end{equation}
where $ b(\Lambda_{\alpha}), c(\Lambda_{\alpha})$ are functions of the (cross)-diffusion coefficients, the competition terms $uv$, and the status $c$~\cite{Nakao2010, Mimar2019, Vidal-Franco2017}. The set of solutions to Eq.~\eqref{eq:charac} for all modes $\alpha$, correspond to a dispersion relation $\lambda\left(\Lambda_{\alpha}\right)$. The Turing instability occurs when at least one of the modes become unstable, indicated by  $Re (\lambda_{\alpha}) >0$ which happens when $c(\Lambda_{\alpha})<0$, and the corresponding mode is denoted $\alpha_c$~\cite{Nakao2010, Mimar2019}. The full details of the calculation are shown in Supplementary Section S1 and Fig. S1. 

After the onset of the Turing instability, the system reaches a steady-state concentration of speakers for each node (city) which is normalized according to $\tilde u = \frac{\langle u \rangle}{\langle u \rangle + \langle v \rangle}$, and similarly for $v$. Here the $\langle \ldots \rangle$ denotes averaging over multiple realizations of the simulation corresponding to different initial conditions. The concentration of speakers at the district level is then a population weighted-sum over all constituent nodes. The full details of the normalization and data aggregation procedure is described in Supplementary Section S2 and Fig. S2. In what is to follow, for the sake of simplicity, we assume that both competing languages (cross)-diffuse at the same rate.

\par

\section{Results}

\subsection{Galicia}
\begin{figure}[t!]
	\centering
	\captionsetup{justification=raggedright, font=small, labelfont=bf}
	
	\includegraphics[width =1.0\columnwidth]{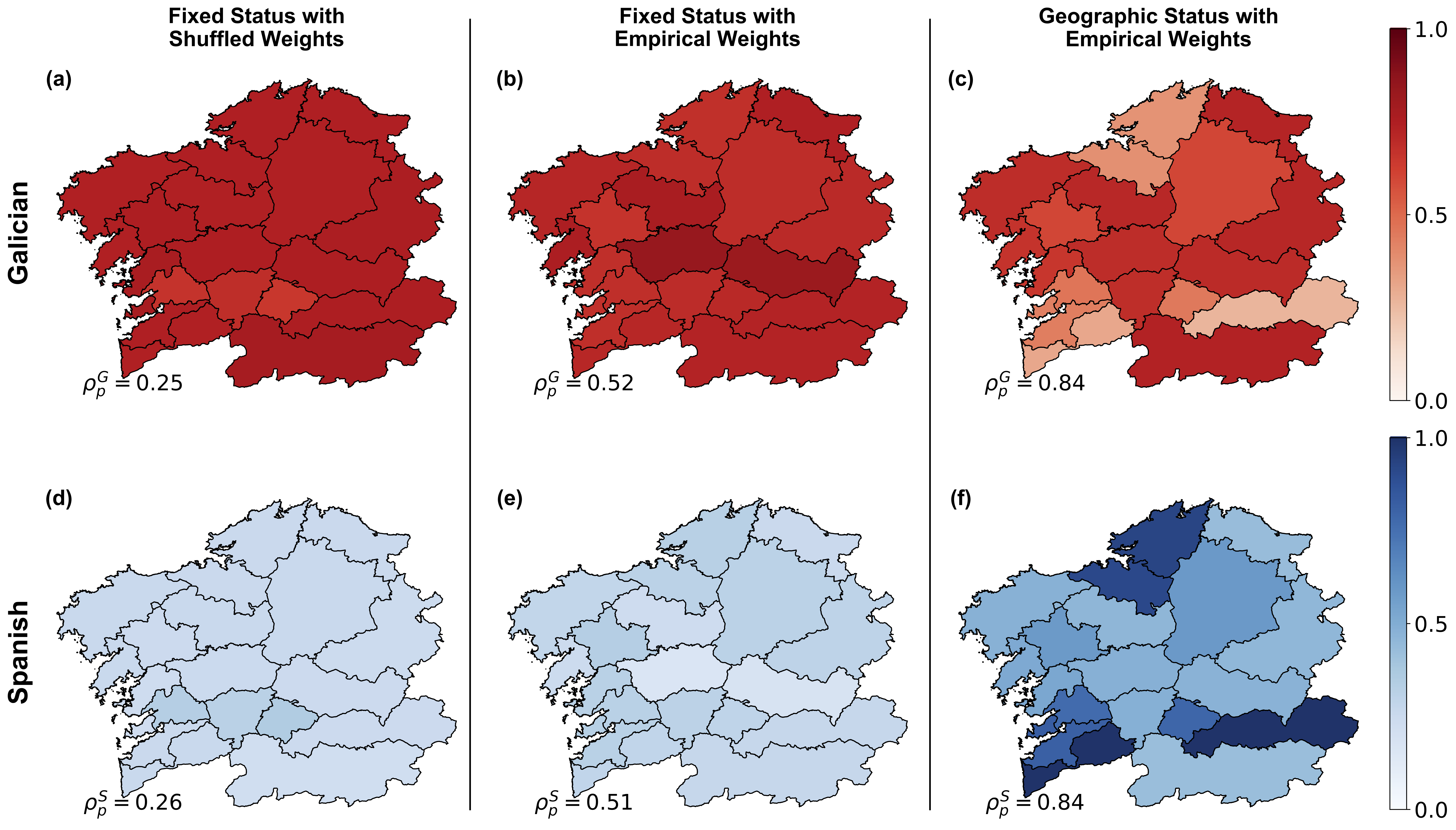}

	\caption{Testing the effect on linguistic prevalence in Galicia for each ingredient in our model. In \textbf{(a), (d)} we test the effect of network topology alone, by assigning equal status to every district ($c=0.5$), and shuffling the weights of the link-strengths according to Fig.~\ref{fig:rawmap}{\bf b}, with the effect of removing the spatial nature of the network. In \textbf{(b), (e)}  we restore the spatial nature of the network, while maintaining a fixed status for every district. Finally, in \textbf{(c), (f)}, we use the spatial network in combination with the empirical geographic distribution of status for each language. Pearson correlation-coefficients in each panel indicate the comparison between the simulated and empirical concentration of speakers.} 
	\label{fig:geodist}  
\end{figure}
We begin our analysis with the case of Galicia. We seek to uncover the role of each underlying mechanism in the observed empirical trends, and therefore systematically probe the effect of the model constituents, starting with the underlying network mediating the population-interactions. Recent results suggest that Turing patterns in networks are influenced and stabilized primarily by network topology provided the distribution of links is heavy-tailed~\cite{Mimar2019}. 

While the mobility networks we consider are spatial, we first check the extent to which the linguistic patterns can be explained solely by the heavy-tailed nature of the network. To do so we randomly assign each node a weight $s$ sampled from the empirical distribution $P(s)$ for Galicia (Fig.~\ref{fig:rawmap}{\bf b}), in effect removing any information about the spatial location of the nodes, and treating the network as a purely topological graph. In addition we set the value of the status $c = 0.5$ everywhere in the region, the diffusion coefficients to $d_u, d_v = 0.01$ and $\gamma = 2.1$ for the cross-diffusion coefficients. These numbers were chosen to drive the system to the onset of the Turing instability. In Fig.~S3{\bf a,d} we show the results of the simulation averaged over 100 realizations of the process, compared to the empirical data, as a scatter plot of the districts according to the preponderance of Galician and Spanish, and in Fig.~\ref{fig:geodist}{\bf a,d} we show the spatial linguistic distributions. 

The scatter plots indicate relatively few nodes differentiate from their initial fixed-points for both Galician and Spanish. The agreement with the empirical data is rather poor with a Pearson correlation-coefficient of $\rho_p ^{G} = 0.25$ for Galician and $\rho_p ^{S} = 0.26$ for Spanish. One can also check the relative prevalence of linguistic speakers in each region by ranking districts by the concentration of speakers for each language, and then compute the rank correlation-coefficient. In Fig. S4{\bf a,d} we show the scatter plot of the simulated and empirical data in terms of the rank of each district. The Spearman correlation coefficient for both Galican and Spanish is $\rho_s ^G = \rho_s^ S = 0.07$. The results indicate that network topology by itself is a poor indicator of the observed linguistic prevalence.

Next, we restore the spatial nature of the network maintaining both $P(s)$ as well as the geographic position of the nodes, i.e links between nodes are established according to Eq.~\eqref{eq:gravity}, and re-run the simulation with the same parameters. We show the results in Fig.~\ref{fig:geodist}{\bf b,e} where we plot the spatial distribution and in Fig.S3{\bf b,e} which shows the scatter plot of concentrations. We find improved correspondence for both Galician ($\rho_p ^G = 0.52$) and Spanish with ($\rho_p ^S = 0.51$), although there is an overestimation of Galician speakers, and a corresponding underestimation of Spanish speakers in about half the districts. This can be explained by the choice of a positive value for $c$  in every district, which biases the result towards favoring Galician speakers. A choice of negative $c$ for each district would reverse the trends. This is also reflected in Fig. S4{\bf b,e} for the rank scatter plots, where we find $\rho_s ^G = \rho_s ^S = 0.63$. Nevertheless, the reasonable agreement between simulation and data for a majority of districts points to an important role played by the geographic networks in linguistic evolution. By itself, however, it is not enough to explain the full picture.

Next, we incorporate the geographical distribution of the status parameter $c$ into our framework. Surveys and polls conducted in Galicia reveal 12 districts where Galician is perceived to have higher status ($c >0$), and 8 districts for the case of Spanish ($c <0$)~\cite{Galicia-fraction}. For those districts where residents report a higher status for Galician we set $c = 0.5$, and for those that prefer Spanish we set $c = -0.5$. We then re-run the simulation for the same set of (cross)-diffusion coefficients as before, and report our results in Fig.~\ref{fig:geodist}{\bf c,f} and Fig.~S3{\bf c,f} for the spatial distribution and scatter plots respectively. We now find significantly better agreement with the empirical data for both types of languages, with $\rho_p ^G = \rho_p ^S = 0.84$. A similar effect is seen in the relative abundance as reflected by the rank scatter plots shown in Fig. S4{\bf c,f} with $\rho_s ^G = \rho_s^S = 0.85$.  Taken together, the results indicate that the heavy-tailed nature of the geographical mobility network coupled with the spatial correlation of the status parameters (along with their asymmetry) is a good predictor of the linguistic prevalence in Galicia.

\subsection{Southern Carinthia}

\begin{figure}[t!]
	\centering
	
	\captionsetup{justification=raggedright, font=small, labelfont=bf}
	\includegraphics[width =1.0\columnwidth]{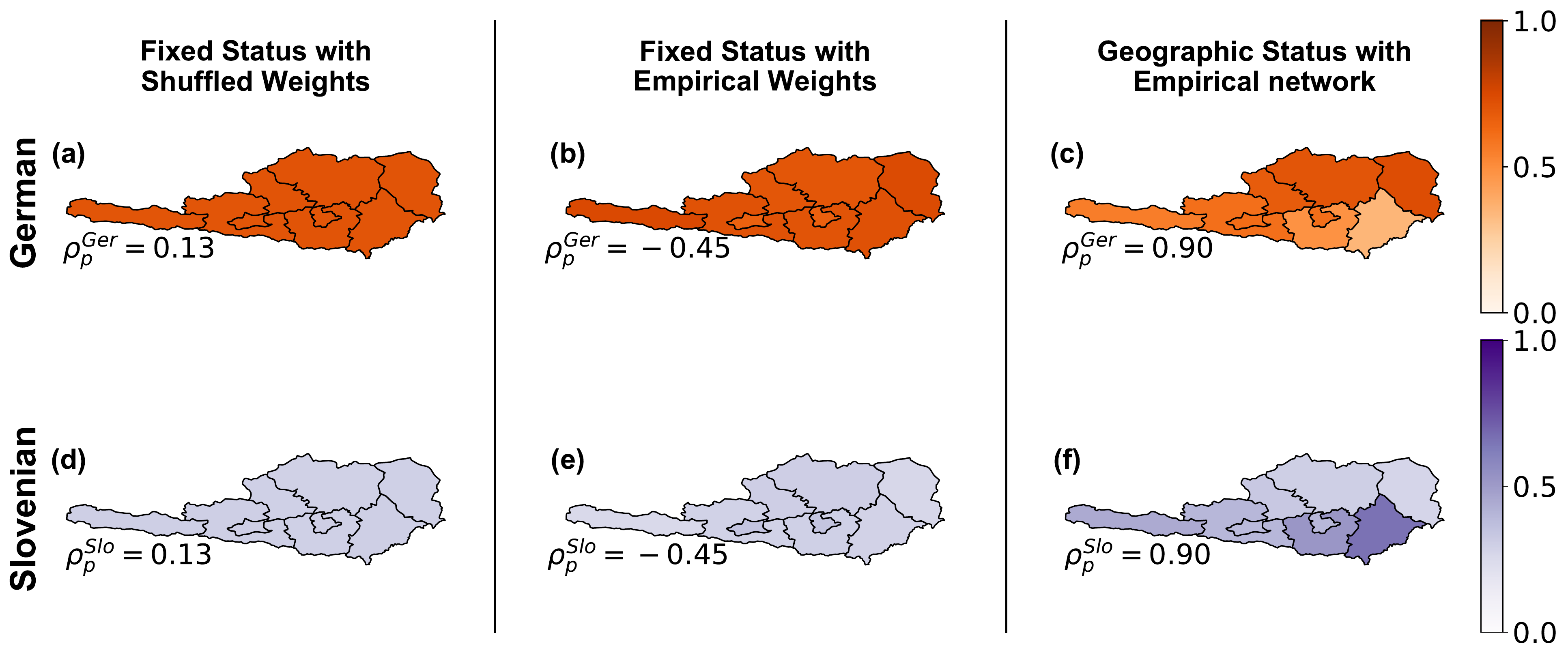}

	\caption{Testing the effect on linguistic prevalence in Southern Carinthia for each ingredient in our model. In \textbf{(a), (d)} we test the effect of network topology alone, by assigning equal status to every district ($c=0.5$), and shuffling the weights of the link-strengths according to Fig.~\ref{fig:rawmap}{\bf b}, with the effect of removing the spatial nature of the network. In \textbf{(b), (e)}  we restore the spatial nature of the network, while maintaining a fixed status for every district. Finally, in \textbf{(c), (f)}, we use the spatial network in combination with the empirical geographic distribution of status for each language. Pearson correlation-coefficients in each panel indicate the comparison between the simulated and empirical concentration of speakers.}
	\label{fig:geodist2}
\end{figure}

To check whether our results are unique to Galica, or generalizable to other regions, we now repeat the analysis for Southern Carinthia. We adjust the (cross)-diffusion constants to generate an instability range coinciding with the eigenvalue distribution of the empirical network Laplacian; now $d_u =  d_v$ = $ 0.1$ and $\gamma = 21$. We once again, start by randomly assigning weights to nodes sampled from the empirical distribution $P(s)$ seen in Fig.~\ref{fig:rawmap}{\bf d}, assign the same value of the status $c= 0.5$ in all districts and simulate the linguistic evolution. Much like in Galica, we find the same poor agreement with the empirical data in terms of both the fraction of speakers ($\rho_p^{Ger} = 0.13, \rho_p^{Slo} = 0.13$, Fig.~\ref{fig:geodist2}{\bf a,d} and S5 {\bf a,d}) as well as their relative abundance $(\rho_s^{Ger} = 0.2, \rho_s^{Slo} = 0.2$, Fig. S6 {\bf a,d}), indicating that here too, the topological nature of the network by itself is a poor predictor of linguistic prevalence. 

Next, we consider the geographical network with weights assigned according to the nodes' positions (Eq.~\eqref{eq:gravity}). The results are shown in Fig.~\ref{fig:geodist2}{\bf b,e} and Fig.~S5{\bf b,e}. Unlike in Galicia, here we find poor correspondence with the data. The Pearson correlation coefficients are now $\rho_p^{Ger} = \rho_p^{Slo} = -0.45$ and the Spearman correlation coefficients are $\rho_s^{Ger} = \rho_s^{Slo} = -0.35$. The negative values for the correlation stem from only a few regions, and it is more accurate to say that there appears to be no correlation between the simulated and empirical data. The contrast with Galicia is striking, and is potentially due to the network in Southern Carinthia being an order of magnitude smaller in size. In this relatively small setting, the geographical mobility network has minimal-to-no-role in predicting the linguistic patterns.

Additionally, we do not have access to how residents perceive the status of each language given that there are no (known) surveys or polls. A reasonable choice in determining the status, however is to infer it from the proportion of speakers. That is, in those regions where German is the majority tongue we assume German has the higher status, and similarly regions with majority Slovenian speakers are assigned a higher status for Slovenian. Correspondingly in German majority districts we set $c= 0.5$ and $c = -0.5$ for Slovenian majority regions. We re-run the simulation for the same set of (cross)-diffusion coefficients as before, and report our results in Fig.~\ref{fig:geodist2}{\bf c,f} and Fig.~S5{\bf c,f} for the spatial distribution and scatter plots respectively. In this case, we find even better agreement with the data as compared to Galicia, with $\rho_p ^{Ger} = \rho_p^{Slo} = 0.9$. A similar trend is seen for the relative abundance (Fig. S6 {\bf c,f}) with $\rho_s ^{Ger} = \rho_s ^{Slo} = 0.9$. Thus, in this case, while the network plays a limited role, accounting for the asymmetry and spatial correlation of the status, the Turing mechanism produces good agreement with the empirical linguistic distributions in in Carinthia.

\section{Discussion}

In this manuscript we have presented a minimal formulation to explain the observed linguistic trends in two regions of Europe where languages co-exist. Our model, based on the Turing mechanism has as its primary ingredients, a reaction-diffusion model where language species spread and retreat in the same fashion as it occurs in predator-prey dynamics, the mobility network between locations based on the gravity model, coupled with the asymmetries and the geographical distribution in how speakers perceive a given language. Unlike in other descriptions of linguistic evolution, the model constituents are set up in a way, such that we can tease out the effects of each component. Another advantage of our framework as compared to existing formulations is the need for minimal empirical input, as well as its generalizability to multiple settings. Given that the language dynamics occurs on a discrete network we are able to simultaneously capture microscopic and macroscopic dynamics without the rise of pathologies such as ``language islands'' due to the lack of diffusive fronts in non-contiguous regions. 

While patterns have been known to be stabilized by heterogenous network topologies in other settings, considering just the network topology by itself without considering its spatial nature, leads to poor agreement with our results and empirical trends. Once one accounts for the spatial location of nodes, the model gets about half the districts right in Galicia. Note that this occurs despite assigning both languages, Galician and Spanish, equal status among residents. In our version, we assigned higher status to Galican and despite this, the model was able to accurately reproduce some of the districts where Spanish is the majority language. This points to strong evidence for the spatial mobility network of contacts playing an important role in language interaction and diffusion. Similar results were seen for the relative abundance of languages, that is, ranking districts based on the concentration of speakers of each language. 

Interestingly enough, the same was not seen for Southern Carinthia, where the spatial network appeared to have little-to-no predictive power in terms of the concentration of German and Slovenian speakers. Nevertheless, when coupled with a bimodal distribution for the status parameter (reflecting asymmetries in how languages are perceived), we got very good agreement in Galicia as well as Southern Carinthia, both in terms of the concentration of speakers and the relative abundance of the languages. Note that, in the latter case, we were able to produce good agreement with the empirical values, despite not knowing the actual values of the status parameters for each language.

Our results are notable, given our minimal set of assumptions as well as little recourse to empirical parameters. Of course, to go beyond this one would need tailored models with more granular data and the introduction of more region-specific parameters. Additionally, we do not consider more complex facets of linguistic prevalence such as bilingualism~\cite{Seoane2017,MussaJuane2019}, however such features can be in principle introduced through an additional term in Eq.~\eqref{eq:lvd}. Nevertheless, our objective here was to focus on uncovering the (potential) basic mechanisms of linguistic evolution and compare it against empirical trends, and not necessarily attempt to reproduce exactly the concentration of speakers in a region. We anticipate our formulation will be quite useful in understanding linguistic prevalence in those regions with scarce data on the relevant parameters.

\section*{Acknowledgments}
SM, GG and JP acknowledge support from the Global Research Network program through the Ministry of Education of the Republic of Korea and the National Research Foundation of Korea (NRF-2016S1A2A2911945). APM and MMJ are supported by the Spanish Ministerio de Econom\'ia y Competitividad and European Regional Development Fund, contract RTI2018-097063-B-I00 AEI/FEDER, UE; by Xunta de Galicia, Research Grant No. 2018-PG082, and the CRETUS Strategic Partnership, AGRUP2015/02, supported by Xunta de Galicia. All these programs are co-funded by FEDER (UE). GG and SM also thank the University of Rochester for financial support.

 \def\eprintprefix{}
 \def\eprint#1{}
\bibliographystyle{naturemag}
\bibliography{refer}

\newpage
\setcounter{figure}{0}    
\phantomsection
\setcounter{equation}{0}  
\setcounter{tocdepth}{2}
\setcounter{section}{0}

\addcontentsline{toc}{chapter}{Supporting Information}
%
%
	%
	%
	\renewcommand{\listfigurename}{Supplementary Figures}
	\renewcommand{\listtablename}{Supplementary Tables}
	\renewcommand{\thesection}{S\arabic{section}}
	\renewcommand{\thefigure}{S\arabic{figure}}
	\renewcommand{\thetable}{S\arabic{table}} 
	\renewcommand{\theequation}{S\arabic{equation}}

\noindent{\LARGE{\bf Supplementary Information}}

	\section{Turing instability}
	
			Expanding the functions $f(u_i,v_i) = c u_i v_i + u_i(1-u_i)$ and $g(u_i,v_i) = -c u_i v_i + v_i(1-v_i)$ to first-order around the fixed points $u_0$, $v_0$ via perturbations $\delta u_i$,$\delta v_i$, Eq. (3) can be written in linearized form as:
	
	\begin{equation}
	\begin{pmatrix} 
	\frac{du_i}{dt}    \\ 
	\frac{dv_i}{dt}  
	\end{pmatrix} =  \begin{pmatrix} 
	f_u & f_v   \\ 
	g_u &  g_v
	\end{pmatrix}
	\begin{pmatrix} 
	u_i -u_0 \\ 
	v_i  -v_0
	\end{pmatrix} + \sum_{j=1}^{N} L_{ij}
	\begin{pmatrix} 
	d_u+ a_{uv}v_0 & a_{uv}u_0   \\ 
	a_{vu} v_0 &  d_v + a_{vu}u_0
	\end{pmatrix} \begin{pmatrix} 
	u_j   \\ 
	v_j  
	\end{pmatrix}, 
	\label{eq:linde}
	\end{equation} 
	with the Jacobian and diffusion matrix are respectively defined as:
	\begin{equation}
	\textbf{J}=J|_{u_0,v_0} = \begin{pmatrix} 
	f_u & f_v   \\ 
	g_u &  g_v
	\end{pmatrix}
	\quad\mathrm{and}\quad 
	\textbf{D}=
	 D|_{u_0,v_0} = \begin{pmatrix} 
	 d_u+ a_{uv}v_0 & a_{uv}u_0   \\ 
	 a_{vu} v_0 &  d_v + a_{vu}u_0
	 \end{pmatrix}=\begin{pmatrix} 
	 D_{uu} & D_{uv}   \\ 
	 D_{vu} &  D_{vv} 
	 \end{pmatrix}.
	\end{equation}
	The eigenvalue equation for the Laplacian matrix is: $\sum_{j=1}^{N} L_{ij} \phi_{j}^{(\alpha)} = \Lambda_{\alpha}\phi_{i}^{(\alpha)},   \alpha = 1 \cdots N$. In terms of small perturbations, Eq.~\eqref{eq:linde} becomes:
	\begin{align}
	\frac{d \delta u_{i}}{dt} = f_u \delta u_i + f_v \delta v_i + \sum_{j=1}^N L_{ij} \textbf{D} \delta u_i, \nonumber \\
	\frac{d  \delta v_{i}}{dt} = g_u \delta u_i + g_v \delta v_i + \sum_{j=1}^N L_{ij} \textbf{D} \delta v_i.
	\label{eq:lindepert}
	\end{align}
	The perturbations can be expanded over the set of Laplacian 
	eigenvectors as $\delta u_i (t) = \sum_{\alpha = 1}^{N} B_u^{(\alpha)} exp[\lambda_{\alpha} t ] \phi_{i}^{(\alpha)}$ and $\delta v_i (t) = \sum_{\alpha = 1}^{N} B_v^{(\alpha)}  exp[\lambda_{\alpha} t ] \phi_{i}^{(\alpha)}$. Substituting these into Eq.~\eqref{eq:lindepert} we obtain the following eigenvalue equation:
	\begin{equation}
	\lambda_{\alpha}
	\begin{pmatrix} 
	B_u^{(\alpha)}   \\ 
	B_v^{(\alpha)}
	\end{pmatrix} = \begin{pmatrix} 
	f_u + D_{uu} \Lambda_{\alpha} & f_v + D_{uv}\Lambda_{\alpha}   \\ 
	g_u + D_{vu} \Lambda_{\alpha} & g_v + D_{vv}\Lambda_{\alpha}
	\end{pmatrix}
	\begin{pmatrix}
	B_u^{(\alpha)} \\
	B_v^{(\alpha)}
	\end{pmatrix}
	\end{equation}
	The characteristic equation of this system is given by:
	\begin{equation}
	\lambda_{\alpha}^2 + b(\Lambda_{\alpha})\lambda_{\alpha} + c(\Lambda_{\alpha}) = 0
	\label{eq:chareq}
	\end{equation}
	where:
	 \[b(\Lambda_{\alpha})= -[Tr(\textbf{J}) + Tr(\textbf{D}) \Lambda_{\alpha}],\]
	\[c(\Lambda_{\alpha}) =  Det(\textbf{D})\Lambda_{\alpha}^2 + [D_{uu}g_{v} + f_{u} D_{vv} - f_{v} D_{vu} -D_{uv}g_u] \Lambda_{\alpha} + Det(\textbf{J}).\]
	The solutions to Eq.~\eqref{eq:chareq} are then: $\lambda_{\alpha_1}  = \frac{-b(\Lambda_{\alpha}) + \sqrt{b(\Lambda_{\alpha})^2 - 4c(\Lambda_{\alpha})}}{2}$
	and $\lambda_{\alpha_2}  = \frac{-b(\Lambda_{\alpha}) - \sqrt{b(\Lambda_{\alpha})^2 - 4c(\Lambda_{\alpha})}}{2}$. When diffusion starts, the only solution with positive real part is $\lambda_{\alpha_1}$. Thus, we define the dispersion relation in terms of Laplacian eigenvalues  as Re($\lambda(\Lambda_{\alpha}))$ with roots $\Lambda_{\alpha_1}$ and $\Lambda_{\alpha_2}$, defining the instability range.
	\par
\begin{figure*}[t!]	
		\centering
		\includegraphics[width=0.85\linewidth]{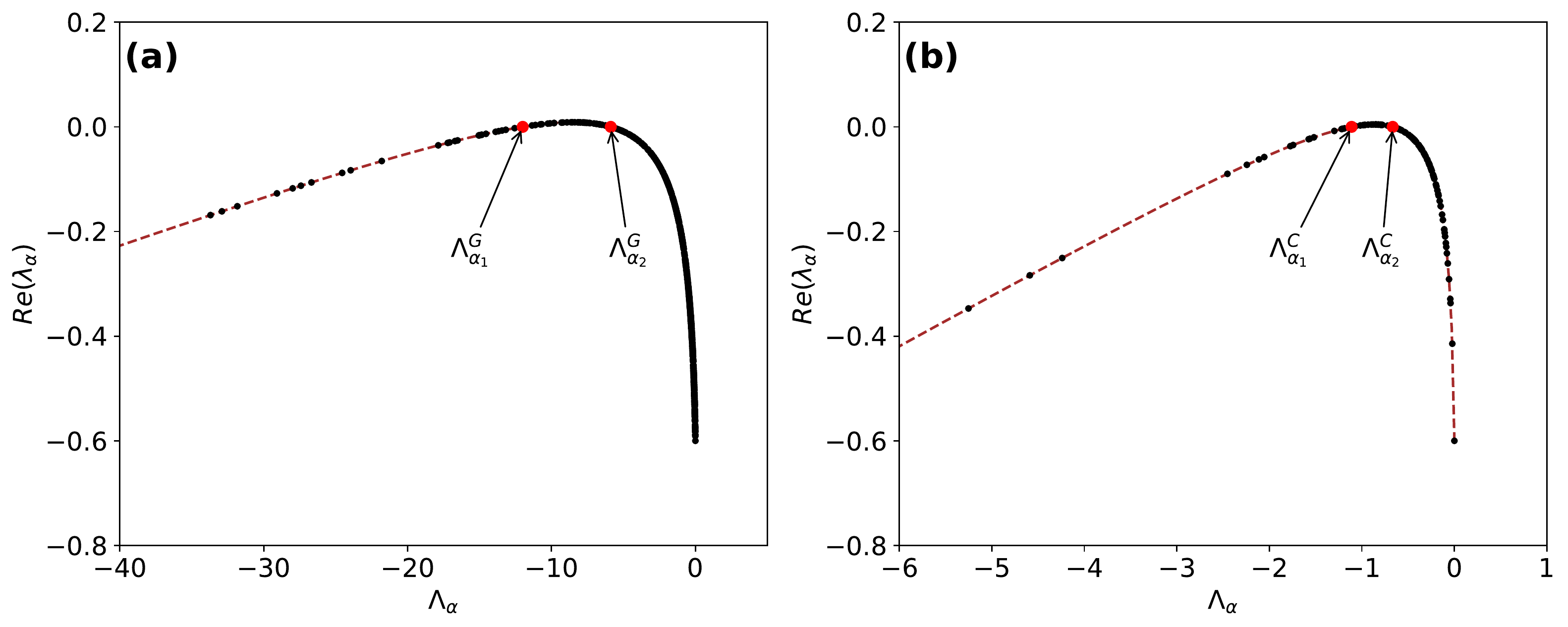}\\
		\caption[ Eigenvalue distributions of empirical networks in instability range]{ Eigenvalue distributions of empirical networks of \textbf{(a)} Galicia and \textbf{(b)} Carinthia along the dispersion curve  Re($\lambda(\Lambda_{\alpha})$). The instability ranges are marked by $[\Lambda_{\alpha_1}^G, \Lambda_{\alpha_2}^G]$ and $[\Lambda_{\alpha_1}^C, \Lambda_{\alpha_2}^C]$ for Galicia and Carinthia respectively.  Differentiation of nodes are triggered by the instable growth factors Re($\lambda(\Lambda_{\alpha})) > 0$ which correspond to eigenvalues overlapping the instability ranges.}
		\label{sifig:eigenvalus}
	\end{figure*}

	The Turing instability is triggered when the eigenvalues $\Lambda(\alpha)$ become unstable, which indicates that the corresponding growth factors in the dispersion relation Re($\lambda(\Lambda_{\alpha})$) become positive. In Fig.~\ref{sifig:eigenvalus} we show the eigenvalues distribution for the geographic networks of Galicia \textbf{(a)} and Carinthia  \textbf{(b)} along the curve Re($\lambda(\Lambda_{\alpha})$). The unstable modes are the eigenvalues that lie in the range $[\Lambda_{\alpha_1}^G, \Lambda_{\alpha_2}^G]$ and $[\Lambda_{\alpha_1}^C, \Lambda_{\alpha_2}^C]$ respectively. 
	\par
	
	\section{Data normalization and aggregation}
	\begin{figure*}[t!]	
		\centering
		\includegraphics[width=0.85\linewidth]{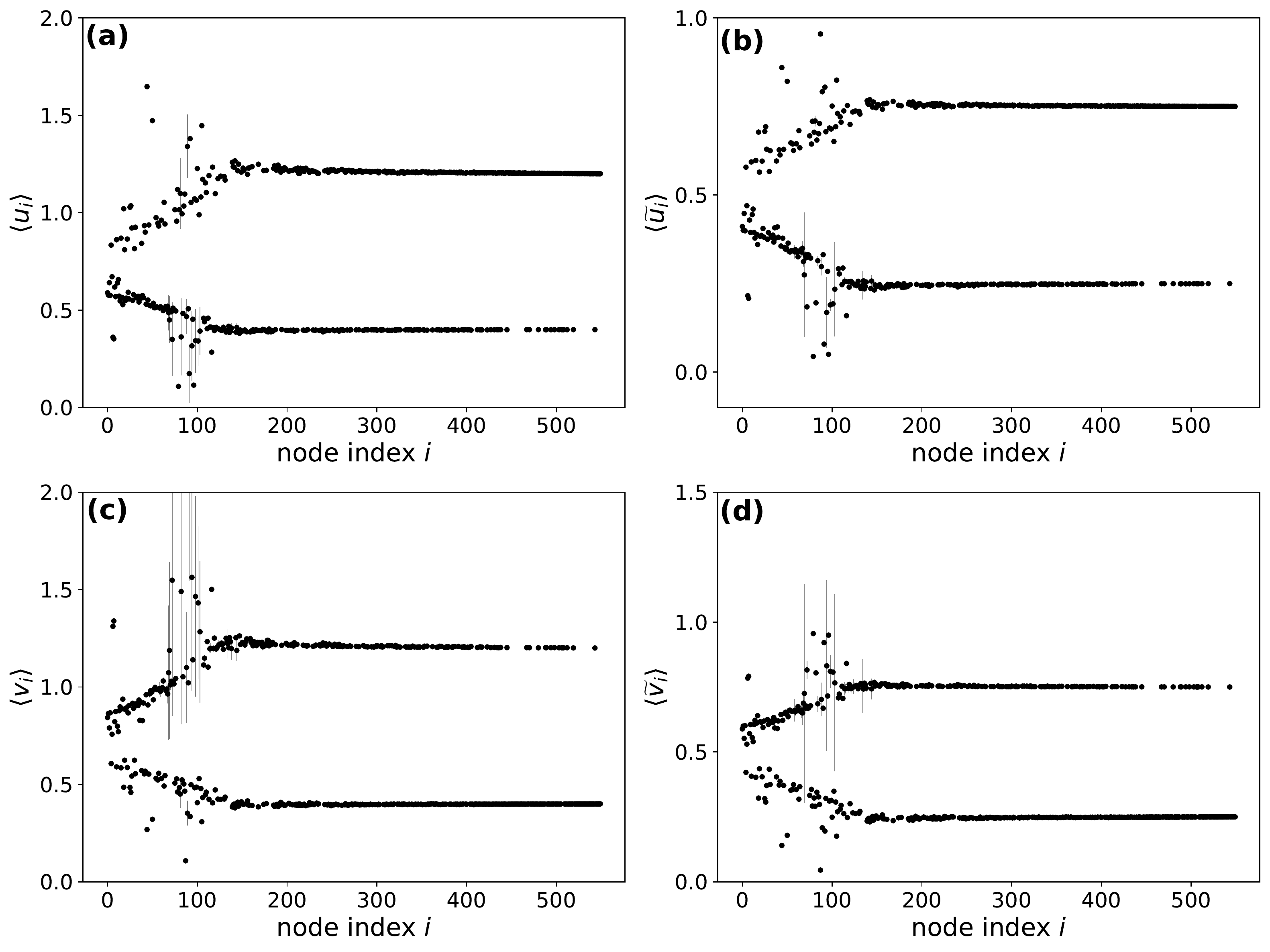}\\
		\caption[Data Normalization]{Averaged simulation results after multiple realization with different random initial perturbations. Panels  \textbf{(a)} and \textbf{(c)} are the concentration of speakers of Galician and Spanish respectively for each node.  Panels  \textbf{(b)} and \textbf{(d)} are the normalized fractions of speakers.}
		\label{sifig:normalization}
	\end{figure*}
	Fig.~\ref{sifig:normalization} shows the results of a typical simulation for the region of Galicia using the set of parameters used to generate Fig. 3 in the main manuscript. Galician and Spanish speakers start with the initial fixed points $u_0 = 1.2$  and $v_0 = 0.4$ in districts where $c=0.5$ and $u_0 = 0.4$  and $v_0 = 1.2$ in districts where $c=-0.5$. The concentration of speakers in the new stationary state is shown in Fig.~\ref{sifig:normalization} \textbf{(a)} and \textbf{(c)}. The results are averaged over multiple realizations with different initial random perturbations to the fixed points. The points correspond to the average over multiple realizations and fluctuations are shown as error bars. In Fig.~\ref{sifig:normalization} \textbf{(b)} and \textbf{(d)} we show the normalized concentrations, where nodes are rescaled with the total average concentration per node ( $  \tilde{u}_0 = \frac{\langle u_i \rangle}{\langle u_i \rangle + \langle v_i \rangle}$ and $ \tilde{v}_0  = \frac{\langle v_i \rangle}{\langle u_i \rangle + \langle v_i \rangle}$). The normalized values of the initial fixed points correspond to $\tilde{u}_0 = 0.75; \tilde{v}_0 = 0.25$. 
	\par
	The results are then aggregated to the level of districts by calculating the weighted average of node concentration by the population $p_i$ of the node that they belong to. In other words, Galician and Spanish speakers of district $j$ is calculated by $\langle {u}^j \rangle = \frac{\sum_{i \in j} \tilde{u}_i p_i}{\sum_{i \in j} p_i}$
	and  $\langle {v}^j \rangle = \frac{\sum_{i \in j} \tilde{v}_i p_i}{\sum_{i \in j} p_i}$. The same procedure is used in Carinthia.

	\begin{figure*}[b!]	
		\centering
		\includegraphics[width=0.85\linewidth]{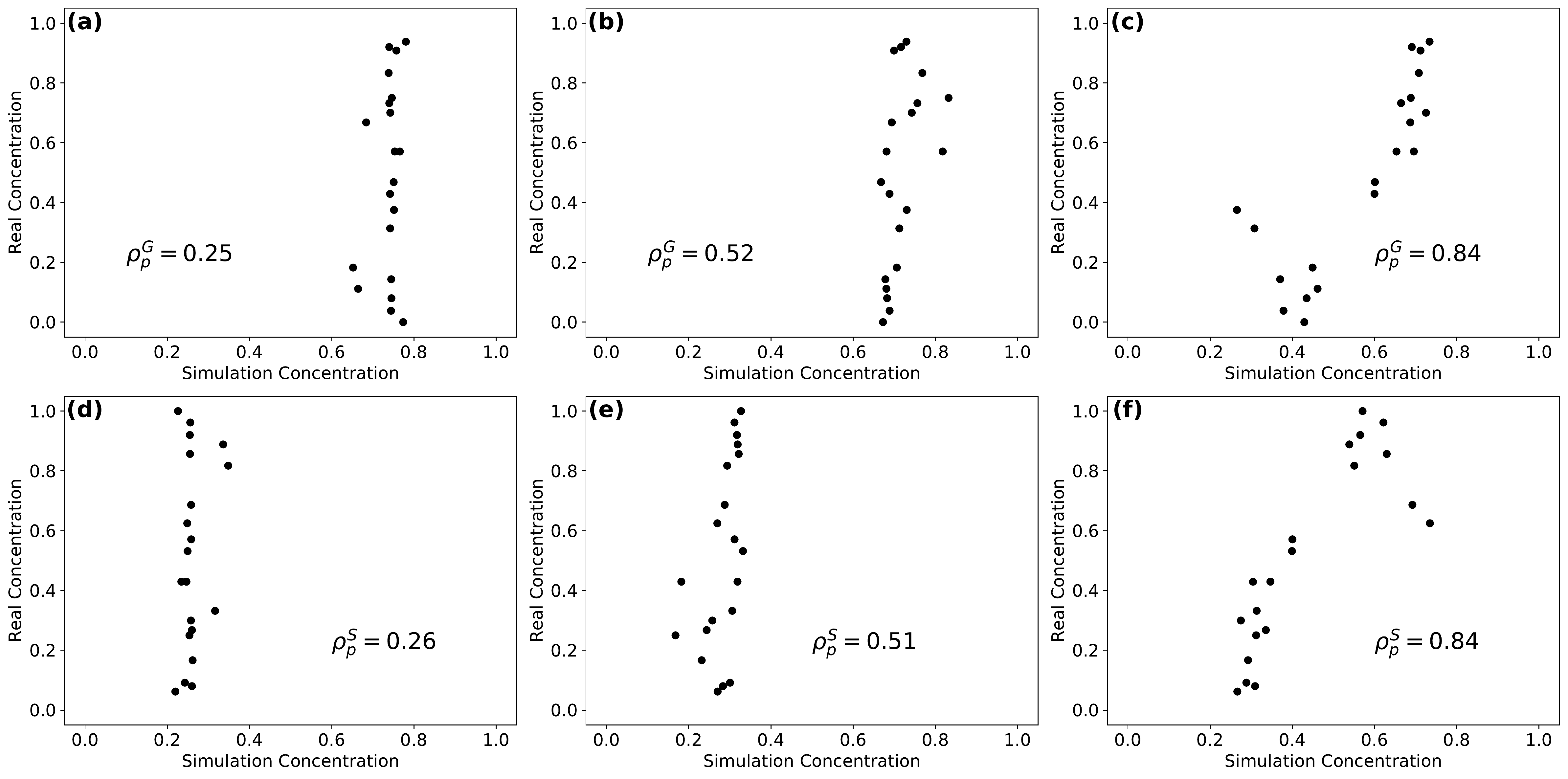}\\
		\caption[Scatter plots for real vs simulation concentrations in Galicia]{Comparison of simulation results with empirical concentrations in Galicia, illustrated in the same order as in Fig.~3\textbf{a, b, c} for Galician and  \textbf{d, e, f} for Spanish speakers. The Pearson correlation-coefficient ($\rho_p$) is reported in each panel.}
	
	\end{figure*}
	
	\begin{figure*}[b!]	
		\centering
		\includegraphics[width=0.85\linewidth]{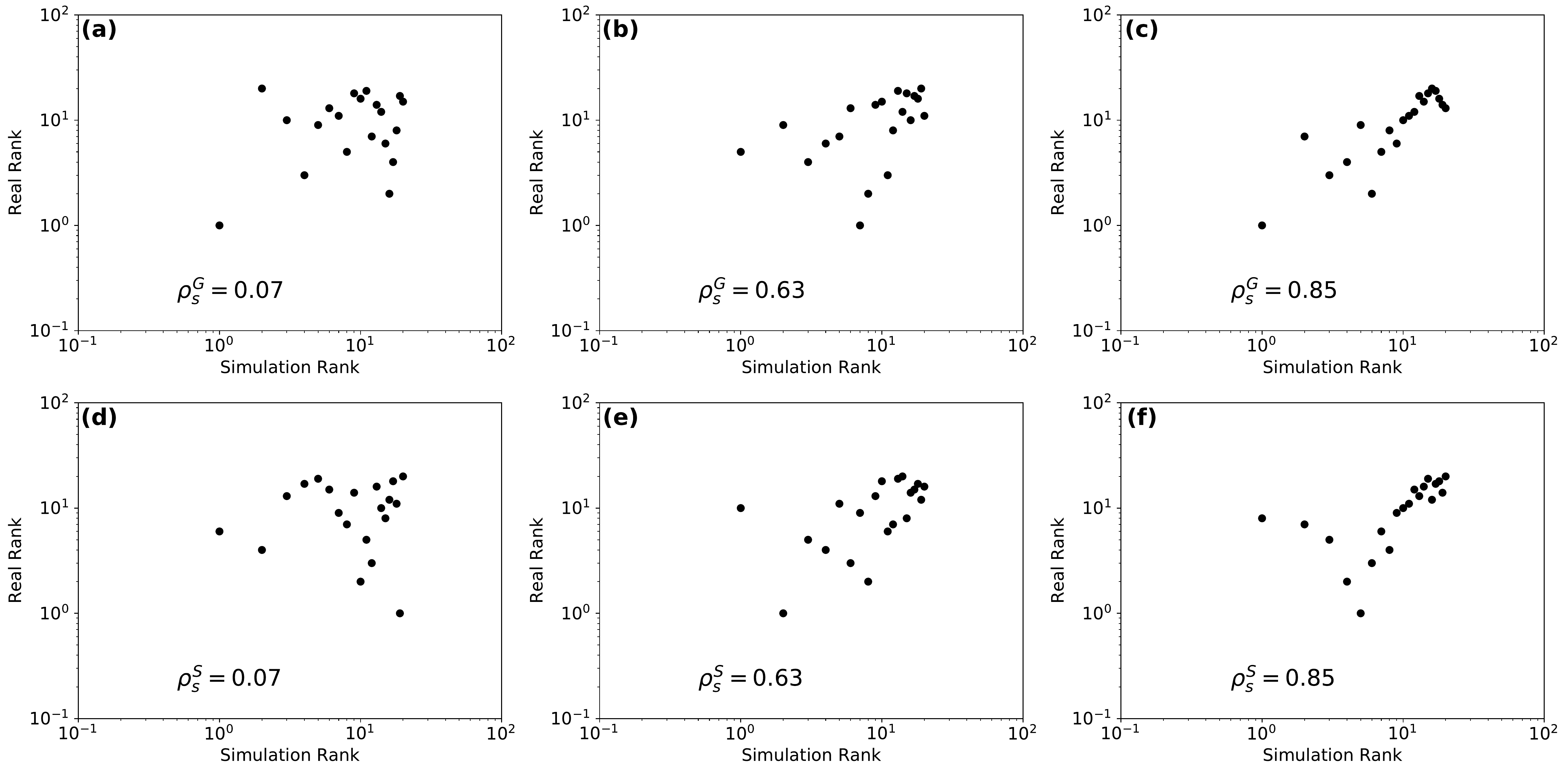}\\
		\caption[Scatter plots for real vs simulation rankings in Galicia]{Comparison of simulation results with empirical rank-ordering of districts in terms of concentrations, in the same order as in Fig.~3\textbf{a, b, c} for Galician and  \textbf{d, e, f} for Spanish speakers. The Spearman correlation-coefficient ($\rho_s$) is reported in each panel.}
	
	\end{figure*}

	\begin{figure*}[t!]	
		\centering
		\includegraphics[width=0.85\linewidth]{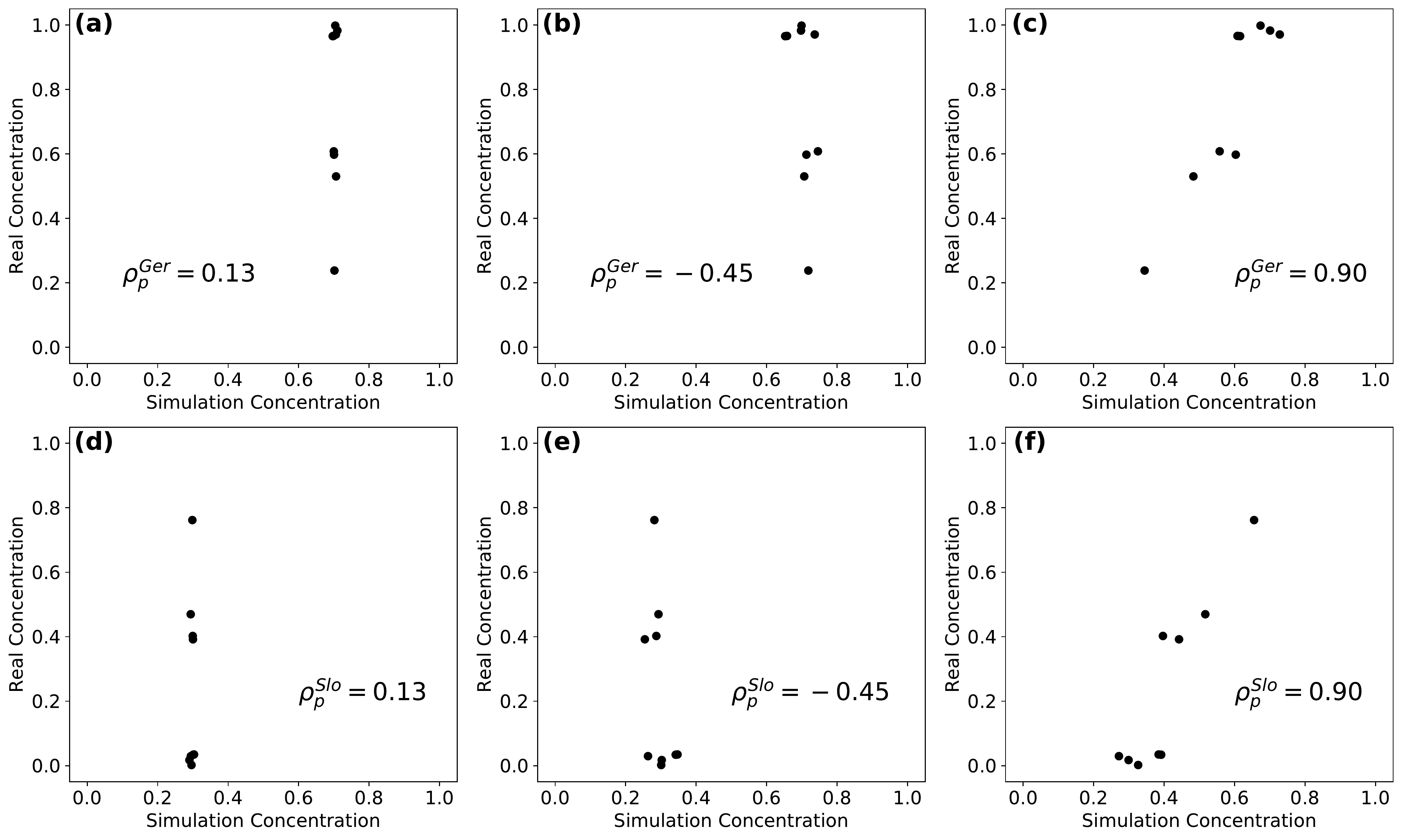}\\
		\caption[Scatter plots for real vs simulation concentrations in Southern Carinthia]{Comparison of simulation results with empirical concentrations in Carinthia, in the same order as in Fig.~4\textbf{a, b, c} for German and  \textbf{d, e, f} for Slovenian speakers. The Pearson correlation-coefficient ($\rho_p$) is reported in each panel.}
		
	\end{figure*}
	
	\begin{figure*}[t!]	
		\centering
		\includegraphics[width=0.85\linewidth]{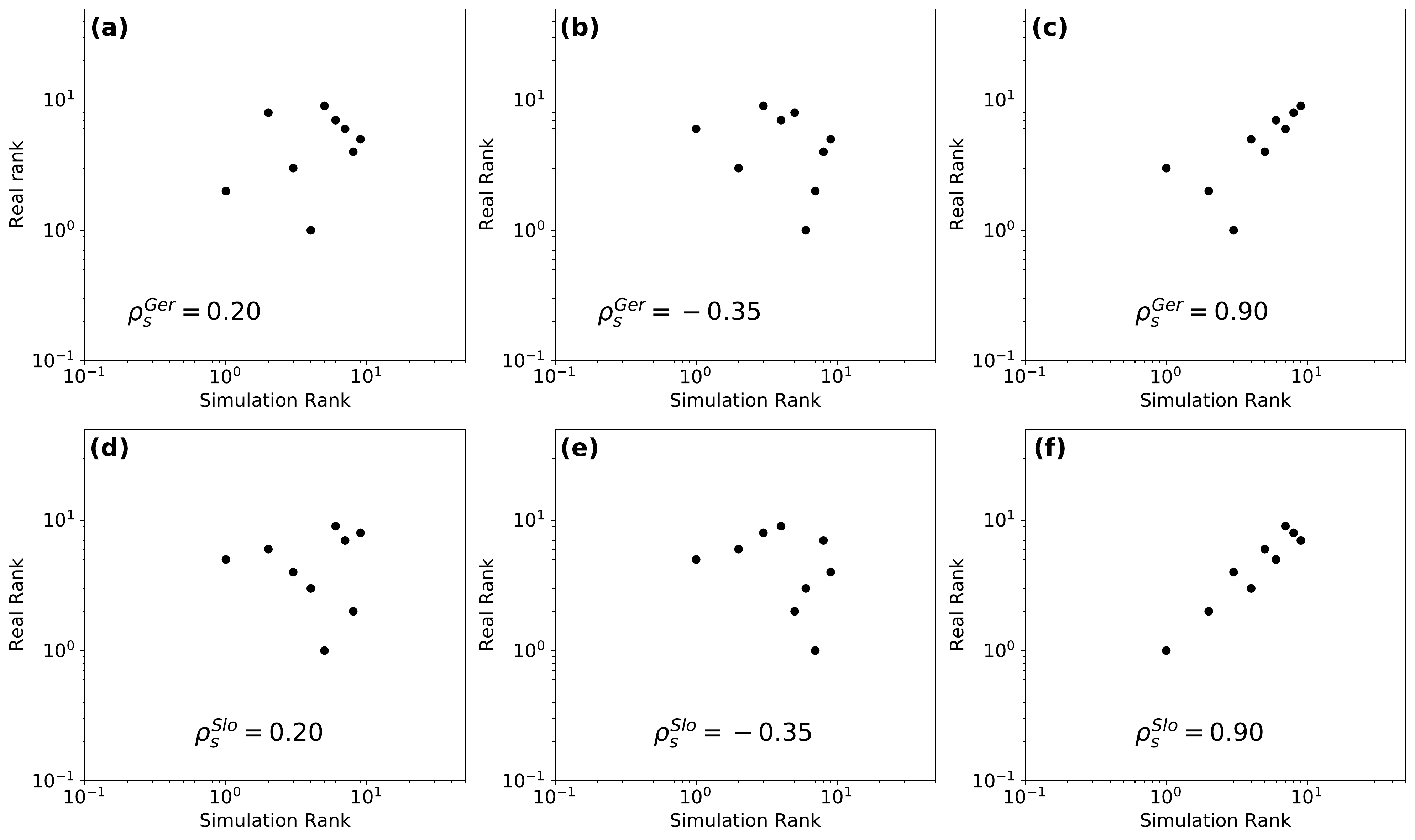}\\
		\caption[Scatter plots for real vs simulation rankings in Southern Carinthia]{Comparison of simulation results with empirical rank-ordering of districts in terms of concentrations, in the same order as in Fig.~4\textbf{a, b, c} for German and  \textbf{d, e, f} for Slovenian speakers. The Spearman correlation-coefficient ($\rho_s$) is reported in each panel.}
		
	\end{figure*}

	%


\end{document}